\documentstyle[aps,prb,preprint] {revtex}
\begin{document}
\draft
%\twocolumn[\hsize\textwidth\columnwidth\hsize\csname %
%@twocolumnfalse\endcsname
\newcommand{\lsvo} {Li$_2$VOSiO$_4$ }
\newcommand{\lsco} {La$_{2-x}$Sr$_x$CuO$_4$ }
\newcommand{\lscotu} {La$_{1.98}$Sr$_{0.02}$CuO$_4$ }
\newcommand{\lscofri} {La$_{1.97}$Sr$_{0.03}$CuO$_4$ }
\newcommand{\la} {$^{139}$La }
\newcommand{\cu} {$^{63}$Cu }
\newcommand{\cuo} {CuO$_2$ }
\newcommand{\ybco} {YBa$_{2}$Cu$_3$O$_{6.1}$ }
\newcommand{\ybcoca} {Y$_{1-x}$Ca$_x$Ba$_2$Cu$_3$O$_{6.1}$ }
\newcommand{\yt} {$^{89}$Y }
\newcommand{\etal} {{\it et al.} }
\newcommand{\ie} {{\it i.e.} }
\hyphenation{a-long}
%
%
%
%

%%XXX \special{src: 33 LSVPRB1.TEX} %Inserted by TeXtelmExtel

\title{
Magnetic and thermodynamic properties of \lsvo: a two-dimensional $S=1/2$ frustrated 
antiferromagnet on a square lattice 
}
\author{R. Melzi, 
S. Aldrovandi, F. Tedoldi and
P. Carretta\footnote{e-mail: carretta@fisicavolta.unipv.it}} 
\address{Dipartimento di Fisica ``A. Volta'' e Unit\'a INFM di Pavia,
Via Bassi 6, 27100 Pavia, Italy} 
\author{
P. Millet}
\address{
Centre d'Elaboration des Mat\'eriaux et d'Etudes Structurales, CNRS, 31055 Toulouse Cedex, 
France
} 
\author{
F. Mila}
\address{
Institut de Physique Th\'eorique, Universit\'e de Lausanne, CH-1015 Lausanne,
Switzerland}
\date{\today}
\maketitle
%%%%%%%%%%%%%%%%%%%%
\widetext
%%%%%%%%%%%%%%%%%%%%
% \vskip -1cm
\begin{abstract}

%%XXX \special{src: 63 LSVPRB1.TEX} %Inserted by TeXtelmExtel

NMR, $\mu$SR, magnetization and specific heat  measurements in \lsvo powders
and single crystals are reported. Specific heat and magnetization measurements evidence
that \lsvo is a frustrated two-dimensional $S=1/2$ Heisenberg antiferromagnet
on a square lattice with a superexchange coupling $J_1$, along the sides of 
the square, almost equal to $J_2$, the one along the diagonal ($J_2/J_1=1.1\pm 0.1$
with $J_2 + J_1 = 8.2\pm 1$ K). At $T_c\simeq 2.8$ K a phase transition to a low temperature
collinear order is observed. $T_c$ and the sublattice magnetization, derived from NMR and
$\mu$SR, were found practically independent
on the magnetic field intensity up to $9$ Tesla. The critical exponent of the sublattice
magnetization was estimated $\beta\simeq 0.235$, nearly coincident with the one predicted for
a two-dimensional XY system on a finite size. The different  magnetic properties
found above and below $T_c$ are associated with the modifications in the spin hamiltonian 
arising from a structural distortion occurring just above $T_c$.

%%XXX \special{src: 79 LSVPRB1.TEX} %Inserted by TeXtelmExtel

%%XXX \special{src: 82 LSVPRB1.TEX} %Inserted by TeXtelmExtel

%%XXX \special{src: 85 LSVPRB1.TEX} %Inserted by TeXtelmExtel

%%%%%%%%%%%%%%%%%%%%
% \leftskip 54.8pt
% \rightskip 54.8pt
%%%%%%%%%%%%%%%%%%%%

%%XXX \special{src: 92 LSVPRB1.TEX} %Inserted by TeXtelmExtel

%%XXX \special{src: 95 LSVPRB1.TEX} %Inserted by TeXtelmExtel

%%XXX \special{src: 98 LSVPRB1.TEX} %Inserted by TeXtelmExtel

%%XXX \special{src: 101 LSVPRB1.TEX} %Inserted by TeXtelmExtel

%%XXX \special{src: 104 LSVPRB1.TEX} %Inserted by TeXtelmExtel

%%XXX \special{src: 107 LSVPRB1.TEX} %Inserted by TeXtelmExtel

%%XXX \special{src: 110 LSVPRB1.TEX} %Inserted by TeXtelmExtel

%%XXX \special{src: 113 LSVPRB1.TEX} %Inserted by TeXtelmExtel

\end{abstract}
\newpage
%%%%%%%%%%%%%%%%%%%%
% \leftskip 54.8pt
% \rightskip 54.8pt
%%%%%%%%%%%%%%%%%%%%
%
%]
%%%%%%%%%%%%%%%%%%%%
\narrowtext
%%%%%%%%%%%%%%%%%%%%
%\twocolumn

%%XXX \special{src: 129 LSVPRB1.TEX} %Inserted by TeXtelmExtel

%%XXX \special{src: 132 LSVPRB1.TEX} %Inserted by TeXtelmExtel

%
%
%	
	\section{Introduction}
\label{intro}

%%XXX \special{src: 140 LSVPRB1.TEX} %Inserted by TeXtelmExtel

In recent years one
has witnessed an extensive investigation of quantum phase transition in low-dimensional
$S=1/2$ Heisenberg antiferromagnets (QHAF) as a function of doping, magnetic field and 
disorder\cite{QHAF}. For example, two-dimensional QHAF (2DQHAF) 
have been widely 
studied in order to evidence   
a phase transition from the renormalized
classical to the quantum disordered regime 
upon charge doping \cite{HTCSC}.  Another possibility to drive quantum 
phase transitions in a 2DQHAF
is to induce a sizeable
frustration. In particular, for a square lattice with an exchange coupling along the
diagonal $J_2$ about half of the one along the sides 
of the square $J_1$ (see Fig. 1a),
a crossover to a spin-liquid ground state
is expected \cite{Chandra1,Schulz,Sorella}. 
For $J_2/J_1\lesssim 0.35$  N\'eel order is realized, while
for $J_2/J_1\gtrsim 0.65$ a collinear order should develop. 
The collinear order (see Fig. 1a), which
 can be considered as formed by two interpenetrating N\'eel 
sublattices with staggered magnetization ${\bf n_1}$ and  ${\bf n_2}$, is characterized
by an Ising order parameter 
$\sigma= {\bf n_1}. {\bf n_2} =\pm 1$  \cite{Chandra2}. The two 
values of $\sigma$ correspond to the two  collinear  configurations 
which can develop at low
temperature (hereafter $T$),
one with spins ferromagnetically
aligned along the $x$ axis, corresponding to a magnetic wave-vector 
${\bf Q}=(0, \pi/a)$,
the other with spins ferromagnetically aligned along the $y$ axis
(${\bf Q}=(\pi/a,0)$). At a certain temperature $T_c$ a phase transition occurs and
the system choses among the $x$ or $y$ collinear configurations.
The precise boundaries of the $J_2/J_1$
phase diagram for a frustrated 2DQHAF are unknown and could be modified 
by the presence of a finite third
neighbour coupling \cite{Chandra2}. Most of these theoretical predictions
have not found  an experimental
support so far, 
mainly due to the absence of systems which can be regarded as prototypes
of frustrated 2DQHAF on a square lattice with $J_2$ close to $J_1$, even if some
frustrated 2DQHAF with different topologies have been recently studied \cite{Kim}. 
Moreover, a theoretical description of the spin dynamics of these frustrated 2DQHAF
as a function of $T$ and magnetic field intensity is still missing. 

%%XXX \special{src: 186 LSVPRB1.TEX} %Inserted by TeXtelmExtel

%%XXX \special{src: 189 LSVPRB1.TEX} %Inserted by TeXtelmExtel

Recently, two vanadates which can be considered as
prototypes of frustrated 2DQHAF on a square lattice have been discovered \cite{Melzi}:
\lsvo and Li$_2$VOGeO$_4$. 
These two isostructural compounds are characterized by a layered structure
containing V$^{4+}$ ($S=1/2$) ions \cite{Millet} (see Fig. 1b). 
The structure of $V^{4+}$ layers 
suggests that the superexchange couplings between 
first and second neighbours are similar. It is however difficult
a priori to decide which one should dominate: first neighbours are 
connected by two superexchange channels, but they are located in pyramids looking 
in opposite directions and are not exactly in the same plane, whereas second neighbours are
connected by one channel, but are located in pyramids looking 
in the same directions and are in the same plane.
On the basis of NMR and susceptibility it has been possible to demonstrate that
in \lsvo $J_2/J_1$ is of the order of unity and that the ground state
is collinear \cite{Melzi}, as expected for $J_2/J_1 \gtrsim 0.65$\cite{Chandra2}.

%%XXX \special{src: 208 LSVPRB1.TEX} %Inserted by TeXtelmExtel

In this paper we present a detailed study of the magnetic and thermodynamic properties
of \lsvo by means of NMR, $\mu$SR, magnetization and specific heat measurements. In particular,
we show that the spin dynamic and static properties above the collinear 
ordering temperature $T_c$ 
are consistent with the ones theoretically predicted for 
a frustrated 2DQHAF with $J_2/J_1\simeq 1$. The phase transition to the collinear phase seems
to be triggered by a structural distortion occuring just above $T_c$, which possibly modifies
the superexchange couplings and lifts the degeneracy among the two ground state
configurations. The critical exponent of the sublattice magnetization and 
the independence of $T_c$ on the magnetic field intensity up to $9$ Tesla, suggest that
the transition is driven by the  XY anisotropy.     

%%XXX \special{src: 222 LSVPRB1.TEX} %Inserted by TeXtelmExtel

The paper is organized as folows: in Section II we present the experimental results 
obtained by each technique; in Section III we discuss the experimental results in the light
of numerical and analytical results for frustrated 2DQHAF on a square lattice, first 
for $T> T_c$ and then for $T< T_c$; the main conclusions are summarized in Section IV.

%%XXX \special{src: 229 LSVPRB1.TEX} %Inserted by TeXtelmExtel

%%XXX \special{src: 232 LSVPRB1.TEX} %Inserted by TeXtelmExtel

%
\section{Experimental Aspects and Experimental  Results}
\label{sec:1}
\subsection{Sample Preparation, Specific Heat and Magnetization}

%%XXX \special{src: 239 LSVPRB1.TEX} %Inserted by TeXtelmExtel

\lsvo was prepared by solid state reaction starting from a stoechiometric
mixture of Li$_2$SiO$_3$, V$_2$O$_3$ and V$_2$O$_5$ 
according to the procedure described in Ref. 9.
The sample was analyzed by X-ray powder  diffraction (XRD) using a Seifert C3000
diffractometer with CuK$\alpha$ radiation and then pressed into a $1$ g pellet 
followed by a short sintering in vacuum at $800$ C for $6$ hours.
Single crystals, of average size $1\times 1\times 0.2$ mm$^3$, 
were obtained from \lsvo powder heated at $1150$ C for $2$ hours,
slowly cooled at a rate of $5$ C/hour down to $1000$ C and then furnace cooled down to
room temperature.

%%XXX \special{src: 252 LSVPRB1.TEX} %Inserted by TeXtelmExtel

%%XXX \special{src: 255 LSVPRB1.TEX} %Inserted by TeXtelmExtel

%%XXX \special{src: 258 LSVPRB1.TEX} %Inserted by TeXtelmExtel

%%XXX \special{src: 261 LSVPRB1.TEX} %Inserted by TeXtelmExtel

%%XXX \special{src: 264 LSVPRB1.TEX} %Inserted by TeXtelmExtel

%%XXX \special{src: 267 LSVPRB1.TEX} %Inserted by TeXtelmExtel

%%XXX \special{src: 270 LSVPRB1.TEX} %Inserted by TeXtelmExtel

Specific heat ($C(T)$) measurements have been performed on a sintered pellet of \lsvo
by using a standard homemade adiabatic calorimeter. The  contribution
of the addenda decreased from about 5\% to below 1\% of the total heat capacity 
on decreasing  $T$ from 25 K to 2.5 K.  
At low $T$ the  specific heat  
shows a broad maximum due to the correlated 
spin excitations and a sharp peak around $2.8$ K (see Fig. 2a) associated 
with a second order phase transition, as can be inferred from the 
non-singular behaviour of the entropy around $2.8$ K. Above 20 K a rapid increase, 
originating from phonon excitations 
is observed (see Fig. 2a). In order to accurately estimate the magnetic
contribution $C^m(T)$ (Fig. 2b) to the specific heat one has to subtract the phonon term 
$C^p(T)$, extrapolated to low $T$. $C^p(T)$ 
was observed to follow a Debye law from 20 to 70 K, namely 
\begin{equation}
C^p(T)= 9Nk_B ({T\over \Theta_D})^3\int_0^{\Theta_D\over T} {x^4 e^x\over (e^x - 1)^2} dx ,
\end{equation}
with $\Theta_D\simeq 280$ K the Debye temperature. 
It must be stressed that below $15$ K $C^p(T)$ is negligible with respect to $C^m(T)$, therefore,
any incorrect extrapolation of $C^p(T)$ to low $T$ will not affect  the
estimate of $C^m(T)$ below $15$ K.

%%XXX \special{src: 294 LSVPRB1.TEX} %Inserted by TeXtelmExtel

Magnetization measurements were carried out with a Quantum Design MPMS-XL7
SQUID magnetometer, both on powders and on single crystals. The $T$ dependence
of the spin susceptibility $\chi=M/H$ is shown in Fig. 3. One observes a high $T$
Curie-like behaviour, a low $T$ maximum around 5 K and a kink at $T_c\simeq 2.8$ K,
the same temperature where a peak in the specific heat is detected. The kink is 
better evidenced if one reports the derivative of the susceptibility $d\chi /dT$ (see the upper 
inset to Fig. 3).
The susceptibility above $15$ K can be appropriately fitted by 
\begin{equation}
\chi (T)= \chi_{VV} + c_{\chi}/(T + \Theta) ,
\end{equation}
where $\Theta$ is the Curie-Weiss temperature, $c_{\chi}=N_A g^2 S(S+1)\mu_B^2 /3 k_B$ 
($g$ the Land\'e factor and $\mu_B$ the Bohr magneton) is the Curie constant and $\chi_{VV}$ 
the Van-Vleck term. The best fit of the data in the $T$ range 10-300 K yields
$\Theta=8.2\pm 1$ K \cite{Nota1}, $c_{\chi}=0.34$ emu K/mole and $\chi_{VV}= 4\times 10^{-4}$ emu/mole. 
We point out that the value of $c_{\chi}$ is in good 
quantitative agreement with what one would expect for an $S=1/2$ paramagnet,
while the absolute value of $\chi_{VV}$ is consistent with a separation between
the $d_{xy}$ ground state and the first excited $t_{2g}$ levels of the order of $0.15$ eV, 
which is typical for V$^{4+}$ in a pyramidal environment \cite{Nota2}. 
Below  $T_c$ the $T$ dependence of the susceptibility for magnetic fields $H\parallel c$
and $H\perp c$ is different, as expected in the 
presence of long range order. In particular,
one observes that while for $H\perp c$ the susceptibility progressively diminishes on
decreasing temperature, for $H\parallel c$ it flattens (see the lower inset to Fig. 3), 
suggesting that V$^{4+}$ magnetic moments lie in the $ab$ plane. 

%%XXX \special{src: 323 LSVPRB1.TEX} %Inserted by TeXtelmExtel

The magnetic field dependence of $T_c$, derived either from the kink  in the susceptibility 
or from
the maximum in $d\chi /dT$, was measured from $0.1$ up to $7$ Tesla, where 
the Zeeman energy $g\mu_B H$ is greater than $k_B\Theta$, and, remarkably,
$T_c$ was not observed to vary by more than $0.07$ K, i.e. less than $0.03 T_c$ (see Fig. 4). 

%%XXX \special{src: 331 LSVPRB1.TEX} %Inserted by TeXtelmExtel

%%XXX \special{src: 334 LSVPRB1.TEX} %Inserted by TeXtelmExtel

%%XXX \special{src: 337 LSVPRB1.TEX} %Inserted by TeXtelmExtel

%%XXX \special{src: 340 LSVPRB1.TEX} %Inserted by TeXtelmExtel

\subsection{$\mu$SR}

%%XXX \special{src: 344 LSVPRB1.TEX} %Inserted by TeXtelmExtel

Zero field (ZF) $\mu$SR measurements have been carried out on \lsvo powders at ISIS pulsed
source, both on EMU and MUSR beamlines, using spin-polarized $29$ MeV/c muons. The
time evolution of the muon polarization is characterized by a constant background, due to the sample holder
and cryostat walls,  and by a fast decay which progressively changes from
exponential (see Fig. 5) to gaussian on increasing temperature, for $T > T_c$. Below 
$T_c$ both oscillating and  non-oscillating components are evident, the second one
with an amplitude about half
of the former one, as usually expected in magnetic powders with equivalent muon sites
\cite{Schenck}. 
It must be mentioned that below $T_c$ around 10\% of the total asymmetry is missing, possibly due
to fast precessing muons which cannot be detected at a pulsed muon source. Summarizing,
below $T_c$ the time evolution of the muon polarization was fitted according to
\begin{equation}
P_{\mu}(t)= A_{back}+ A_1e^{-\sigma t} \cos(\gamma B_{\mu} t + \phi) + A_2e^{-\lambda t} ,
\end{equation}
where $A_{back}$ is the sample holder background, $A_1$ is the amplitude of the oscillating
component, with $\gamma= 2\pi \times 135.5$ MHz/Tesla the $\mu^+$ gyromagnetic ratio and
$B_{\mu}$ the local field at the $\mu^+$, while $A_2$ is the ampltude of the non-oscillating
component with $\lambda$ the longitudinal decay rate. Above $T_c$ the polarization was fitted 
by
\begin{equation}
P_{\mu}(t)= A_{back}+ A e^{-\lambda t} e^{-\sigma_N^2 t^2/2}
\end{equation}
where the exponential term is the relaxation induced by the progressive slowing 
down of the V$^{4+}$ spin fluctuations on decreasing temperature, while the gaussian term 
should originate from nuclear dipolar interaction. In particular,  it is likely that  $\mu^+$
localizes close to the apical oxygens, where it is coupled to $^7$Li
nuclear magnetic moments. The gaussian relaxation rate was estimated $\sigma_N=0.34 \pm 0.01$
$\mu s^{-1}$, for $T_c\leq T\leq 4.2$ K, a value typical for relaxation driven by nuclear dipole
interaction \cite{Schenck}. 

%%XXX \special{src: 377 LSVPRB1.TEX} %Inserted by TeXtelmExtel

The $T$ dependence of the local field at the muon $B_{\mu}$ and of the longitudinal
relaxation rate $\lambda$, derived after the fit of the data with Eqs. 3 and 4,
are reported in Figs. 6 and 7, 
respectively.
$B_{\mu}(T)$, which yields the $T$ dependence of  V$^{4+}$ average magnetic
moment, is characterized by a sharp but continuous decrease on approaching $T_c$, while
$\lambda(T)$ is characterized by a divergence at $T_c$, as expected for a second order
phase transition.

%%XXX \special{src: 388 LSVPRB1.TEX} %Inserted by TeXtelmExtel

%%XXX \special{src: 391 LSVPRB1.TEX} %Inserted by TeXtelmExtel

\subsection{$^7$Li and $^{29}$Si NMR}

%%XXX \special{src: 395 LSVPRB1.TEX} %Inserted by TeXtelmExtel

$^7$Li ($I=3/2$) NMR measurements have been carried out both on single crystals and powders,
while $^{29}$Si ($I=1/2$) NMR, due to the reduced sensitivity could be performed only in powder
samples. The measurements have been performed using standard NMR pulse sequences. In particular,
the spectra have been recorded by Fourier transform of half of the echo signal when
the line was completely irradiated or by summing spectra recorded at different frequencies
when it was only partially irradiated. 
The NMR resonance frequency of $^7$Li  was observed to shift to high frequencies
on decreasing temperature, with a trend identical to the one of the macroscopic susceptibility 
(Fig.3). In fact, for $^7$Li NMR shift one can write 
\begin{equation}
^7\Delta {\cal K} (T) = {\sum_j {\cal A}_j \chi(T)\over g\mu_B N_A} + \delta
\end{equation}
where ${\cal A}_j$ is the hyperfine coupling tensor with the $j$-th V$^{4+}$
and $\delta$ the chemical shift.
A $T$ dependent shift was observed both in the single crystals and in the powders,
evidencing a sizeable transferred hyperfine interaction of V$^{4+}$ spins with $^7$Li 
nuclei. From the plot of the shift versus the susceptibility (Fig. 8) the hyperfine coupling 
constants and the chemical shift were determined. It turns out that $\delta= -70\pm 30$ ppm
and that the hyperfine field at $^7$Li is given by
\begin{equation}
\vec h= \sum_j ({\cal A}_{dip})_j \vec S_j + \sum_{i= 1,2}A_t \vec S_i
\end{equation}
where ${\cal A}_{dip}$ is the dipolar coupling with V$^{4+}$ ions, while $A_t=850$ Gauss is the
transferred coupling, which is supposed to arise from the two V$^{4+}$ nearest neighbours only.
On the other hand, $^{29}$Si NMR resonance frequency in the powders is 
constant from room temperature
down to 4 K, pointing out that the hyperfine coupling is of purely dipolar origin in this case.

%%XXX \special{src: 425 LSVPRB1.TEX} %Inserted by TeXtelmExtel

Below $T_c$, in the single crystals, for $H\parallel c$,
$^7$Li NMR spectrum splits in three lines (see Fig. 9): a central one with 
an intensity
about twice of that of two equally spaced satellites. 
The two satellites correspond to $^7$Li sites
with hyperfine fields of equal intensity but opposite orientations, while the central
line corresponds to $^7$Li sites where the hyperfine field cancels out \cite{Melzi}. The $T$
dependence of the satellites shift  is proportional to the amplitude
of V$^{4+}$ magnetic moment and, therefore, it is another method, besides ZF $\mu$SR, 
to determine the temperature
dependence of the sublattice magnetization (see Fig. 10).

%%XXX \special{src: 439 LSVPRB1.TEX} %Inserted by TeXtelmExtel

$^{29}$Si NMR powder spectrum shows a quite different behaviour
at low T. Around $3$ K, still above $T_c$, one observes the appearence
of a shifted narrow peak (see Fig. 11). On decreasing $T$ the low-frequency peak
progressively disappears, while the intensity of the high frequency one increases. This jump in $^{29}$Si NMR shift has to be associated
either with a modification of the chemical shift or of the
hyperfine coupling, suggesting the occurrence
of a structural distortion just above $T_c$. It has to be noticed that, on the contrary,
no anomaly was detected in $^7$Li spectra around $3$ K.
Below $T_c$ $^{29}$Si NMR linewidth is very close to the one above $T_c$, indicating that
the local field at $^{29}$Si site is zero.
 
%%XXX \special{src: 452 LSVPRB1.TEX} %Inserted by TeXtelmExtel

%%XXX \special{src: 455 LSVPRB1.TEX} %Inserted by TeXtelmExtel

Nuclear spin-lattice relaxation rate $1/T_1$ was measured
by exciting the nuclear magnetization either with a  comb of saturating pulses or with
a $180^o$ pulse (inversion recovery sequence). 
Both for $^7$Li and $^{29}$Si the recovery of nuclear 
magnetization towards equilibrium was a single exponential, indicating that for $^7$Li also
the $\pm 3/2 \rightarrow \pm 1/2$ lines were sizeably irradiated during the measurements. 
In fact, at room temperature one can discern the $\pm 3/2 \rightarrow \pm 1/2$ lines 
shifted by $\simeq 40$ kHz from the $+ 1/2 \rightarrow - 1/2$ \cite{Nota3}.
The $T$ dependence of $^7$Li $1/T_1$ is shown in Fig. 12. One observes that $1/T_1$
is constant from room temperature down to $\simeq 3.2$ K, then shows a peak at $T_c$
and rapidly decreases in the ordered phase.  $^{29}$Si $1/T_1$ shows a similar T
dependence below $4.2$ K (Fig. 13), its absolute value, however, is about two orders of magnitude
smaller, supporting the conclusion in favour of a hyperfine coupling of purely dipolar origin.

%%XXX \special{src: 471 LSVPRB1.TEX} %Inserted by TeXtelmExtel

%%XXX \special{src: 474 LSVPRB1.TEX} %Inserted by TeXtelmExtel

%%XXX \special{src: 477 LSVPRB1.TEX} %Inserted by TeXtelmExtel

\section{Discussion}

%%XXX \special{src: 481 LSVPRB1.TEX} %Inserted by TeXtelmExtel

\subsection{Above $T_c$}

%%XXX \special{src: 485 LSVPRB1.TEX} %Inserted by TeXtelmExtel

The $T$ dependence of the susceptibility and of the specific heat allows 
to derive 
information on the basic parameters of the electron spin hamiltonian, namely the
coupling constants and their ratio $J_2/J_1$. For a non-frustrated $S=1/2$ 2D Heisenberg AF
on a square lattice the Curie-Weiss temperature $\Theta= J_1$ nearly coincides with 
the temperature 
where the susceptibility displays a maximum and one has $T_m^{\chi}= 0.935 \Theta$ \cite{Troyer}. 
On the other hand,
in  \lsvo $\Theta= J_2 + J_1 = 8.2 \pm 1$ K is significantly larger than 
$T_m^{\chi}=5.35$ K (see Fig. 3), as expected for a frustrated system. By comparing the measured ratio 
$T_m^{\chi}/\Theta= 0.65\pm 0.07$ with  exact diagonalization and
quantum Monte Carlo (QMC) results it is possible to estimate $J_2/J_1$ \cite{Melzi}. It turns out
that $J_2/J_1$ is close either to $0.25$ or $2.5$ \cite{Melzi}, 
however, it is not possible to say which
of the two coupling constants is larger. We remark that these two values were 
estimated by assuming
$\Theta= 8.2$ K, however, taking into account the uncertainty of $\pm 1$ K in
the estimate of Curie-Weiss temperature and that exact diagonalizations 
provide useful estimates for $J_2/J_1 < 0.4$, while QMC simulations only for $J_2/J_1\gtrsim 2$
\cite{Melzi}, it is difficult to assign an
error bar to these estimates of $J_2/J_1$. 

%%XXX \special{src: 509 LSVPRB1.TEX} %Inserted by TeXtelmExtel

A more accurate determination of the ratio $J_2/J_1$ can be done by analyzing $C^m(T)$ data
in the light of diagonalization results by Singh and Narayanan \cite{Singh} 
and of the numerical
calculations by Bacci et al. \cite{Bacci}. From the numerical results reported in Refs. 15 and 16
it is possible to plot the
amplitude of the specific heat at the maximum $C^m(T_m^C)$ as a function of the ratio $J_2/J_1$
(Fig. 14a).
It turns out that the value $C^m(T_m^C)= (0.436\pm 0.004) R$ found for \lsvo ($R= N_A k_B$)
(see Fig. 2)
is compatible only with $J_2/J_1=0.44\pm 0.01$ or $1.1\pm 0.1$ (Fig. 14a). 
To discriminate among the two ratios one can analyze how $T_m^C= 3.5\pm 0.1$ K
varies as a function of $J_2/J_1$. Since 
\begin{equation}
{T_m^C\over J_1}= {T_m^C\over \Theta}(1+{J_2\over J_1}) ,
\end{equation}   
with $T_m^C/\Theta=0.42 \pm 0.04$, one can check which value of $J_2/J_1$ is compatible
with the  results of $T_m^C/ J_1$ vs. $J_2/J_1$ reported by Singh and Narayanan
\cite{Singh} 
(see Fig. 14b). One observes that Eq. 7 is satisfied only for $J_2/J_1$ around $0.1$ or $1.1$.
Therefore, the only solution which is compatible with the experimental values both of $T_m^C$ and 
$C(T_m^C)$ is $J_2/J_1=1.1\pm 0.1$. This also indicates that $\Theta$ is close to $9$ K (see Fig.
14b) and that $T_m^{\chi}/\Theta\simeq 0.59$. Now, by assuming this value for  
$T_m^{\chi}/\Theta$ one would derive from the analysis of the 
susceptiblity a value $J_2/J_1\lesssim 2$, in agreement with the specific heat analysis,
even if an accurate estimate
with QMC is prevented, since in this range of $J_2/J_1$ the 
results start to suffer from the minus sign problem.
A value of $J_2/J_1$ around $1.1$ also implies that \lsvo lies on the right hand side of the phase diagram
reported in Fig. 1a, where the ground state is expected to be a collinear phase, in 
complete agreement with NMR results below $T_c$ (see later on).

%%XXX \special{src: 542 LSVPRB1.TEX} %Inserted by TeXtelmExtel

Further information on the superexchange constants can be achieved from the analysis of $^7$Li
$1/T_1$.
In the limit $T\gg J_1 + J_2$, $1/T_1$ is constant (see Fig. 12) and, by resorting to
the usual gaussian form for the decay of the spin correlation function, one can write \cite{Moriya}
\begin{equation}
(1/T_1)_{\infty}={\gamma^2\over 2}{S(S+1)\over 3}{\sqrt{2\pi}\over \omega_E}\times
\sum_{k,i,j}  \vert A_{ij}^k\vert ^2 
\end{equation}  
with $A_{ij}^k$ ($i,j= x,y,z$) 
the components of the hyperfine tensor due to the $k^{th}$ V$^{4+}$ and $\gamma$ the nuclear
gyromagnetic ratio.
$\omega_E={\sqrt{J_1^2 + J_2^2} (k_B/\hbar)}\sqrt{2 z S(S+1)/3}$ is the Heisenberg exchange
frequency, where $z=4$ is the number of nearest neighbour spins of a V$^{4+}$ coupled either
through $J_1$ or $J_2$.  By using in Eq. 8 $(1/T_1)_{\infty}=0.2$
ms$^{-1}$, i.e. $^7$Li relaxation rate in the $T$ range $300$ $-$ $3.2$ K, one finds
$\sqrt{J_1^2 + J_2^2}\simeq 8.7$ K, close to what one would derive from susceptibility
and specific heat measurements.

%%XXX \special{src: 562 LSVPRB1.TEX} %Inserted by TeXtelmExtel

On decreasing temperature $^7$Li and $^{29}$Si $1/T_1$ remain constant down to $3.2$ K,
at variance with $\mu^+$ $1/T_1$ (usually called $\lambda$), which
diverges on decreasing T, already at $4.2$ K, due to the growth of the AF correlations.
Both for  nuclei and $\mu^+$ the spin-lattice relaxation is induced by
the fluctuations of the effective local field, driven by the correlated 
spin dynamics, and
one can write
\begin{equation}
1/T_1\equiv\lambda={\gamma^2\over 2N}\sum_{\vec q,\alpha}\vert A_{\vec q}\vert^2 S_{\alpha\alpha}(\vec q ,\omega_R)
\end{equation}
where $\vert A_{\vec q}\vert^2$ is the hyperfine form factor and  
$S_{\alpha\alpha}(\vec q ,\omega_R)$ ($\alpha= x,y,z$) are the components of the dynamical
structure factor at the NMR or $\mu$SR resonance frequency.
One immediately realizes that a different trend of NMR and $\mu$SR $1/T_1$ can
originate from the different form factors, which couple each one
of these probes in a different way with the spin excitations at the critical wave-vector.
In particular, one might suspect that $^7$Li and $^{29}$Si form factors filter out
the AF correlated spin excitations. However, if one considers that $^7$Li is coupled
via a transferred hyperfine interaction  with V$^{4+}$ nearest neighbours (see Eq. 6), one finds
that $^7$Li form factor is little $\vec q$-dependent and that no filtering of the
AF excitations can be envisaged. Moreover, the divergence of $^7$Li and $^{29}$Si 
$1/T_1$ at $T_c$ evidences that the fluctuations at the critical wave vectors cannot
be completely filtered out.

%%XXX \special{src: 588 LSVPRB1.TEX} %Inserted by TeXtelmExtel

Another relevant difference is still present between $\mu$SR and NMR 
measurements. While the former were performed in zero field, NMR $1/T_1$ measurements were
carried out in magnetic fields ranging from $1.8$ to $9$ Tesla, at which the Zeeman energy
is comparable to the superexchange couplings. Therefore, it is tempting to
associate the different behaviour of NMR and $\mu$SR spin-lattice relaxation rates above $T_c$
to a crossover of regime induced by the magnetic field. In particular, the $T$ independent
$T_1$ measured in NMR would be consistent with a quantum critical regime \cite{Memo}, while the exponential 
divergence of
$1/T_1$ ($\lambda$) measured with $\mu$SR would be consistent with a renormalized classical regime
\cite{Memo}, 
where, by
resorting to classical scaling arguments for 2D systems, one can write \cite{CFTD}
\begin{equation}
1/T_1(T)\equiv\lambda(T)\propto \xi (T) =0.493 a \times 
e^{2\pi\rho_s/T} \biggl[ 1- 0.43{T\over J} + O({T\over J})^2\biggr]
\end{equation}
with $\xi$ the in-plane correlation length, $a$ the lattice step and $\rho_s$ the spin stiffness.
From the $T$ dependence of $\lambda$ (see Fig. 7) above $T_c$ one derives $2\pi\rho_s= 7.4$ K, 
less than the value $1.15\Theta$ expected for a non-frustrated 
system \cite{Schultz}. 

%%XXX \special{src: 611 LSVPRB1.TEX} %Inserted by TeXtelmExtel

%%XXX \special{src: 614 LSVPRB1.TEX} %Inserted by TeXtelmExtel

\subsection{Below $T_c$}

%%XXX \special{src: 618 LSVPRB1.TEX} %Inserted by TeXtelmExtel

Since V$^{4+}$ magnetic moments lie in the $ab$ plane, as suggested by susceptibility
measurements (Fig. 3) and by the EPR analysis of the $g$ tensor \cite{Stepanov}, and provided
that the dipolar magnetic field cancels at $^{29}$Si site, one realizes that the order must be
collinear with a critical wave vector ${\bf Q}=(\pi/a , 0)$, 
where $x$ is the direction of the magnetic moments \cite{Melzi}.

%%XXX \special{src: 626 LSVPRB1.TEX} %Inserted by TeXtelmExtel

The second order transition to the low $T$ collinear phase is evidenced
by the peaks in $1/T_1$ and in $d\chi /dT$. It is remarkable to observe that $T_c$
is practically field independent up to at least $9$ Tesla (see Fig. 4), where 
$g\mu_B H/k_B > J_1 + J_2$, while a decrease is expected, with  $T_c$ vanishing for
$g\mu_B H\simeq 6 k_B J_2$, if $J_2/J_1\simeq 1$, i.e. at $H\simeq 20$ Tesla \cite{Jongh}. 
A possible explanation for this peculiar behaviour is that the structural distortion
occuring just above $T_c$, deduced
from $^{29}$Si NMR spectra,  causes an increase in the coupling
constants and that even at $9$ Tesla $g\mu_B H/k_B < J_1 + J_2$.
Another possibility is that \lsvo is a 2D XY system with $T_c$ close to the
corresponding Berezinskii-Kosterlitz-Thouless transition \cite{Jongh}. 

%%XXX \special{src: 640 LSVPRB1.TEX} %Inserted by TeXtelmExtel

%%XXX \special{src: 643 LSVPRB1.TEX} %Inserted by TeXtelmExtel

%%XXX \special{src: 646 LSVPRB1.TEX} %Inserted by TeXtelmExtel

%%XXX \special{src: 649 LSVPRB1.TEX} %Inserted by TeXtelmExtel

Also the $T$ dependence of the sublattice magnetization, derived either from
the local field at the muon or from the splitting of $^7$Li NMR line, was found independent
on the magnetic field intensity from zero up to $9$ Tesla. From ZF$\mu$SR measurements it has
been possible to derive a critical exponent 
$\beta=0.235\pm 0.009$ for the sublattice magnetization (see Fig. 6). 
Remarkably, this value of $\beta$ is very close
to the one predicted for a 2D XY model on a finite size \cite{Bram}. Although some
in-plane anisotropy
can be discerned from the susceptibility data just above $T_c$ (see the inset to Fig. 3),
there is no 
evidence of a crossover from Heisenberg to XY in the $T$ dependence of the correlation
length, derived from $\lambda (T)$ above $T_c$ \cite{Borsa}. 
It is also interesting to observe that
the sublattice magnetization measured by means of $\mu$SR shows a slight high $T$ tail,
as expected in a finite size system \cite{Bram}.
If the order is purely 2D, without long range order along the $c$ axis, one would expect 
$^7$Li nuclei, which lie between V$^{4+}$ layers, to be characterized by a broad 
powder-like NMR spectrum. This is certainly not the case for $T_c- T\gtrsim 0.2$ K (see Fig. 9),
however one cannot exclude from the NMR measurements that the order is 2D 
in the very vicinity of $T_c$.
In fact, since the strong in-plane XY correlations enhance the 3D coupling 
the difference between the 2DXY and 3D ordering temperatures is expected to 
be small, of the order of  the interlayer coupling $J_{\perp}$ \cite{Ding}. An upper limit
for $J_{\perp}$ can be estimated by assuming that $T_c$ is the 3D ordering temperature of a
Heisenberg AF, where $T_c\simeq 0.4 J_{\perp} \xi^2(T_c)$
\cite{Johnst}. From the
temperature dependence of $\mu^+$ relaxation rate $\lambda$ (see Eq. 10) one finds
$\xi (T_c)/a\simeq 5.3$, leading to $J_{\perp}\simeq 0.2$ K. This value is possibly
overestimated and
difficult to justify if one considers the chemical bondings in \lsvo structure.
Therefore, a purely 2D order should be observable only for $T_c-T\lesssim 0.2$ K. 

%%XXX \special{src: 683 LSVPRB1.TEX} %Inserted by TeXtelmExtel

%%XXX \special{src: 686 LSVPRB1.TEX} %Inserted by TeXtelmExtel

%%XXX \special{src: 689 LSVPRB1.TEX} %Inserted by TeXtelmExtel

Although the nature of this phase transition remains to be clarified, one can argue
that the insensitivity both of the N\'eel temperature and of the critical exponent 
of the sublattice magnetization
to the magnetic field indicates that the phase transition is driven by the XY anisotropy.

%%XXX \special{src: 696 LSVPRB1.TEX} %Inserted by TeXtelmExtel

In 3D magnets with two or more possible pitch vectors ${\bf Q}$, 
the ordering usually corresponds to a choice of pitch vector.
The situation is often more complicated at lower temperature,
and further transitions corresponding to other combinations
of the pitch vectors or to the appearance of higher harmonics
have been reported. Besides, the 
relevant parameter for the nature of the transition is the product
$N=n\times m$, where $n$ is
the number of components of the order parameter (3 for Heisenberg)
and $m$ is the number of equivalent wave-vectors\cite{jensen}. 
As a consequence, the resulting transition can have a large critical 
exponent $\beta$, typically around 0.4, or might in some cases be discontinuous.

%%XXX \special{src: 711 LSVPRB1.TEX} %Inserted by TeXtelmExtel

The results reported in the present paper suggest that the transition is
split into two transitions: First a structural transition, as revealed by
Si NMR, then an ordering transition, as seen at the Li site. A natural 
question arises as to whether the Ising degree of freedom corresponding to
the two possible collinear states is associated with the structural distortion
or with the magnetic ordering. We believe that the first possibility is the 
most likely both on experimental and theoretical grounds. Experimentally, the 
small value of the exponent $\beta$ is typical of layered magnets with XY
symmetry. If the parameter $N$  was
increased by a factor 2 with respect to the number
of components of the order parameter due to the Ising degree of freedom, 
one would
not expect to observe such a small exponent. 
Besides, the choice of a pitch vector
for the collinear phase renders the two directions inequivalent, and this is likely to 
be coupled to the lattice and to be associated with a structural distortion.

%%XXX \special{src: 730 LSVPRB1.TEX} %Inserted by TeXtelmExtel

One has to notice that the structural distortion occuring just above 
$T_c$ may have modified the spin hamiltonian. Therefore, a discussion of the properties 
of the ordered phase on the basis of the parameters extracted above $T_c$ could be misleading.
Nevertheless, one has to notice that, to be consistent with a collinear order, $J_2/J_1$
must be larger than $\simeq 0.65$ also below $T_c$.

%%XXX \special{src: 738 LSVPRB1.TEX} %Inserted by TeXtelmExtel

Information on the coupling constants below $T_c$ can be derived from the 
$T$ dependence of $^7$Li
nuclear spin-lattice relaxation. Below $T_c$ $^7$Li $1/T_1$ 
is mainly driven by two-magnon Raman processes  \cite{Pincus}, leading to
a $T^3$ $T$ dependence for $T\gg\Delta$, the gap in the spin-wave spectrum,
and to $1/T1\propto T^2 exp(-\Delta/T)$ for $T\ll\Delta$. The low $T$ dependence of 
$^7$Li $1/T_1$ turns out to be activated and, by fitting the data for $T\leq 2.2$ K
with the latter expression, 
one finds $\Delta = 6\pm 1$ K \cite{Nota4}.  This value of the spin-wave gap is quite large if compared
to the value of $\Theta= J_1 + J_2$, estimated from susceptibility measurements above $T_c$, and
would imply an axial anisotropy $D\simeq\Theta= 8.2\pm 1$ K ($D\sim\Delta^2/(J_1+J_2)$),
which is quite large for
V$^{4+}$. In fact, the values of the $g$ factor estimated from ESR measurements are very close
to $2$ and yield a value of $D< 1$ K \cite{Stepanov}. Moreover, if $D\simeq\Theta$ 
\lsvo should behave as an Ising system, not as
an XY or Heisenberg one, in sharp contrast to the experimental findings. Thus, one is tempted
to argue that 
the low $T$ collinear phase is characterized by coupling constants slightly larger than the
ones determined above $T_c$, so that $D\ll J_1 + J_2$  and its absolute value is smaller.

%%XXX \special{src: 760 LSVPRB1.TEX} %Inserted by TeXtelmExtel

Finally, one has to expect that frustration also causes a reduction of the staggered
magnetization due to the enhancement of quantum fluctuations. 
The $T\rightarrow 0$ average magnetic moment of V$^{4+}$ ions 
can be obtained from $^7$Li NMR spectra below $T_c$. By extrapolatimg to $T\rightarrow 0$
the splitting of $^7$Li NMR satellites and taking into account the hyperfine couplings given
 by Eq. 6, one can estimate a V$^{4+}$ magnetic moment $\mu(T\rightarrow 0)\simeq 0.24 \mu_B$.
This value is reduced not only with respect to the value $0.65 \mu_B$ expected for
a non-frustrated 2DQHAF, but also with respect to the value derived numerically by Schulz et al.
\cite{Schultz} for $J_2/J_1\simeq 1$, suggesting that probably below $T_c$ $J_2/J_1\lesssim 1$.   

%%XXX \special{src: 772 LSVPRB1.TEX} %Inserted by TeXtelmExtel

%%XXX \special{src: 775 LSVPRB1.TEX} %Inserted by TeXtelmExtel

%%XXX \special{src: 778 LSVPRB1.TEX} %Inserted by TeXtelmExtel

%%XXX \special{src: 781 LSVPRB1.TEX} %Inserted by TeXtelmExtel

\section{Conclusion}

%%XXX \special{src: 785 LSVPRB1.TEX} %Inserted by TeXtelmExtel

In conclusion, it has been shown that \lsvo is a prototype of a frustrated 2DQHAF on a squre lattice
with $J_2/J_1\simeq 1.1$ and $J_2 + J_1= 8.2\pm 1$ K. Its ground state is a collinear phase,
as expected for $J_2/J_1\gtrsim 0.65$. The phase diagram as a function of the magnetic
field intensity is characterized by a constant $T_c(H)$, for $0\leq H\leq 9$ Tesla. 
This observation,
together with the fact that the critical exponent of the magnetization $\beta\simeq 0.235$,
suggest that the transition to the collinear phase is driven by the XY anisotropy. 
The
structural distortion occurring just above $T_c$, is expected to lift 
the degeneracy between the two collinear ground states and to modify
the superexchange couplings. In order to gain further insights on the nature of the
phase transition and on the effective coupling constants below $T_c$ further measurements
with other techniques (e.g. inelastic neutron scattering) are required.

%%XXX \special{src: 801 LSVPRB1.TEX} %Inserted by TeXtelmExtel

%%XXX \special{src: 804 LSVPRB1.TEX} %Inserted by TeXtelmExtel

%%XXX \special{src: 807 LSVPRB1.TEX} %Inserted by TeXtelmExtel

%%XXX \special{src: 810 LSVPRB1.TEX} %Inserted by TeXtelmExtel

\section{Acknowledgments}
A. Lascialfari and J. S. Lord are acknowledged for their help during the SQUID 
and $\mu$SR measurements, respectively. M. C. Mozzati is thanked for the crystal field
calculations. Fruitful discussions
 with L. Capriotti, A. Rigamonti, S. Sorella and R. Vaia are gratefully acknowledged.

%%XXX \special{src: 818 LSVPRB1.TEX} %Inserted by TeXtelmExtel

%%XXX \special{src: 821 LSVPRB1.TEX} %Inserted by TeXtelmExtel

%%XXX \special{src: 1007 LSVPRB1.TEX} %Inserted by TeXtelmExtel

%%XXX \special{src: 1010 LSVPRB1.TEX} %Inserted by TeXtelmExtel

%%XXX \special{src: 1013 LSVPRB1.TEX} %Inserted by TeXtelmExtel

%%XXX \special{src: 1016 LSVPRB1.TEX} %Inserted by TeXtelmExtel

%%XXX \special{src: 1019 LSVPRB1.TEX} %Inserted by TeXtelmExtel

%%XXX \special{src: 1022 LSVPRB1.TEX} %Inserted by TeXtelmExtel

%%XXX \special{src: 1025 LSVPRB1.TEX} %Inserted by TeXtelmExtel

%%XXX \special{src: 1028 LSVPRB1.TEX} %Inserted by TeXtelmExtel

%************FIGURE 1*****************
\begin{figure}
%\vspace{6.5cm}
\caption{a) Schematic phase diagram of a frustrated 2DQHAF on a square lattice 
as a function of the ratio
$J_2/J_1$ of the superexchange couplings. b) Structure of \lsvo projected along [001].
SiO$_4$ tetrahedra are in gray, VO$_5$ pyramids are in black while the gray circles indicate
Li$^+$ position. For details see Ref. 9.}
\end{figure}

%%XXX \special{src: 1040 LSVPRB1.TEX} %Inserted by TeXtelmExtel

%************FIGURE 2*****************
\begin{figure}
%\vspace{6.5cm}
\caption{a) Temperature dependence of \lsvo molar specific heat below $70$ K. The solid lines shows
the phonon contribution to $C(T)$, according to Eq. 1 in the text, with $\Theta_D= 280$ K. b) 
Magnetic contribution to the specific heat, obtained after subtracting the phonon
term corresponding to the solid line in a)}
\end{figure}

%%XXX \special{src: 1051 LSVPRB1.TEX} %Inserted by TeXtelmExtel

%************FIGURE 3*****************
\begin{figure}
%\vspace{6.5cm}
\caption{Temperature dependence of the susceptibility $\chi= M/H$, for $H= 3$ kGauss, in
\lsvo powders. The dashed line shows the best fit according to Eq. 2, for $15\leq T\leq 300$ K.
In the upper inset the derivative $d\chi /dT$ is reported, evidencing
a phase transition around $2.8$ K. In the lower inset magnetization measurements in a
\lsvo single crystal, both for $H$ parallel and perpendicular to the $c$ axis are reported.
The intensity of $M$ for $\vec H\perp c$ have been rescaled for the sake of
comparison.}
\end{figure}

%%XXX \special{src: 1065 LSVPRB1.TEX} %Inserted by TeXtelmExtel

%************FIGURE 4*****************
\begin{figure}
%\vspace{6.5cm}
\caption{Magnetic field versus $T$ phase diagram for \lsvo. The circles indicate
the field dependence of $T_c$ derived from the kink in the susceptibility and/or from
the peak in $d\chi /dT$ (see Fig. 3), while the squares the corresponding values
of $T_c$ determined from $^7$Li NMR spectra (see Fig. 10).}
\end{figure}

%%XXX \special{src: 1076 LSVPRB1.TEX} %Inserted by TeXtelmExtel

%************FIGURE 5*****************
\begin{figure}
%\vspace{6.5cm}
\caption{Time evolution of $\mu^+$ polarization in \lsvo powders for $T$ close to $T_c$.
The solid regular line shows the best fit according to Eq.  3 in the text. The T
stability was within $\pm 5\times 10^{-3}$ K.}
\end{figure}

%%XXX \special{src: 1086 LSVPRB1.TEX} %Inserted by TeXtelmExtel

%************FIGURE 6*****************
\begin{figure}
%\vspace{6.5cm}
\caption{Temperature dependence of the local field at the muon in \lsvo powders, derived
from ZF$\mu$SR measurements. The dashed line indicates the critical behaviour for a
critical exponent of the magnetization $\beta= 0.235\pm 0.009$ (see text).}
\end{figure}

%%XXX \special{src: 1096 LSVPRB1.TEX} %Inserted by TeXtelmExtel

%************FIGURE 7*****************
\begin{figure}
%\vspace{6.5cm}
\caption{Temperature dependence of the muon longitudinal relaxation rate in \lsvo powders.
The solid line indicates the $T$ dependence of $\lambda$ according to Eq. 10,
with a spin stiffness $\rho_s=7.4/2\pi$ K.}
\end{figure}

%%XXX \special{src: 1106 LSVPRB1.TEX} %Inserted by TeXtelmExtel

%************FIGURE 8*****************
\begin{figure}
%\vspace{6.5cm}
\caption{Plot of $^7$Li NMR paramagnetic shift versus the macroscopic susceptibility in \lsvo, for
$H\parallel c$. The solid line shows the best fit yielding a total hyperfine coupling of 
$2.6$ kGauss and a chemical shift $\delta=-70\pm 30$ ppm, for
$\vec H \parallel c$.}
\end{figure}

%%XXX \special{src: 1117 LSVPRB1.TEX} %Inserted by TeXtelmExtel

%************FIGURE 9*****************
\begin{figure}
%\vspace{6.5cm}
\caption{ $^7$Li NMR spectra for $H=1.8$
Tesla along the $c$ axis in a \lsvo single crystal, in the proximity of 
$T_c$.}
\end{figure}

%%XXX \special{src: 1127 LSVPRB1.TEX} %Inserted by TeXtelmExtel

%%XXX \special{src: 1130 LSVPRB1.TEX} %Inserted by TeXtelmExtel

%************FIGURE 10*****************
\begin{figure}
%\vspace{6.5cm}
\caption{ Temperature dependence of the splitting of $^7$Li NMR satellites, 
for $H=1.8$ Tesla along the $c$ axis. The solid line shows the critical
behaviour for an exponent $\beta= 0.24$.}
\end{figure}

%%XXX \special{src: 1140 LSVPRB1.TEX} %Inserted by TeXtelmExtel

%************FIGURE 11*****************
\begin{figure}
%\vspace{6.5cm}
\caption{ $^{29}$Si NMR powder spectrum in \lsvo for $H=1.8$ Tesla in the
proximity of $T_c$. The dotted lines mark the position
of the peak at high and at low $T$.}
\end{figure}

%%XXX \special{src: 1150 LSVPRB1.TEX} %Inserted by TeXtelmExtel

%************FIGURE 12*****************
\begin{figure}
%\vspace{6.5cm}
\caption{ $^7$Li nuclear spin-lattice relaxation rate $1/T_1$ for $\vec H\parallel c$ in
\lsvo, for $H=1.8$ Tesla (open squares) and $9$ Tesla (closed circles). The dotted line
gives the best fit according to the expression for 2-magnon relaxation processes (see text), 
yielding $\Delta= 6\pm 1$ K. In the inset the
corresponding $T$ dependence in the range $1.6$ to $100$ K is reported.}
\end{figure}

%%XXX \special{src: 1162 LSVPRB1.TEX} %Inserted by TeXtelmExtel

%************FIGURE 13*****************
\begin{figure}
%\vspace{6.5cm}
\caption{ $^{29}$Si NMR $1/T_1$ in \lsvo for $H=1.8$ Tesla, for $T\leq 4.2$ K.}
\end{figure}

%%XXX \special{src: 1170 LSVPRB1.TEX} %Inserted by TeXtelmExtel

%************FIGURE 14*****************
\begin{figure}
%\vspace{6.5cm}
\caption{a) Amplitude of the maximum in the molar  specific heat for a 
frustrated 2DQHAF versus $J_2/J_1$.
The open squares represent the data derived from Ref. 15, while the closed circles derived from
Ref. 16. The gray region around $C^m(T_m^C)/R=0.436$ represents the experimental value for this
quantity, inclusive of the error bar. b) $T_m^C/J_1$ (see text) versus $J_2/J_1$ derived from
Ref. 15. The solid lines show the behaviour according to Eq. 7, for values of $\Theta$ 
corresponding to the lower and upper limits of the Curie-Weiss temperature estimated
from susceptibility measurements.}
\end{figure}

%%XXX \special{src: 1185 LSVPRB1.TEX} %Inserted by TeXtelmExtel

%%XXX \special{src: 1188 LSVPRB1.TEX} %Inserted by TeXtelmExtel

%%XXX \special{src: 1191 LSVPRB1.TEX} %Inserted by TeXtelmExtel

%%XXX \special{src: 1194 LSVPRB1.TEX} %Inserted by TeXtelmExtel

%%XXX \special{src: 1197 LSVPRB1.TEX} %Inserted by TeXtelmExtel

%%XXX \special{src: 1200 LSVPRB1.TEX} %Inserted by TeXtelmExtel

%%XXX \special{src: 1203 LSVPRB1.TEX} %Inserted by TeXtelmExtel

%%XXX \special{src: 1206 LSVPRB1.TEX} %Inserted by TeXtelmExtel


\begin{references}
\bibitem{Nota3} On the basis of a point charge approximation one derives an electric
field gradient tensor at $^7$Li nuclei with its main axis tilted by $\simeq 40$ degrees
from the $c$ axis and a quadrupolar frequency of $\simeq 140$ kHz, yielding a shift
of $\pm 1/2 \rightarrow \pm 3/2$ resonance frequency from the central resonance
of $\simeq 60$ kHz, close to the one experimentally observed.
\bibitem{Nota4} This value of $\Delta$ is reduced with respect to the value ($\Delta\simeq 18$ K)
previously reported in Ref. 8. The reason is that in Ref. 8 $1/T_1$ has been fitted with a
simple Arrhenius law, without the $T^2$ prefactor, and over a $T$ range limited to higher 
temperatures.
\end{references}
\end{document}